\documentclass[sn-mathphys]{sn-jnl}

\jyear{2021}%

\theoremstyle{thmstyleone}%
\theoremstyle{thmstyletwo}%

\theoremstyle{thmstylethree}%

\begin{document}

\title[Article Title]{Confidence Interval Construction for Multivariate time series using Long Short Term Memory Network}


\author[1]{\fnm{Aryan} \sur{Bhambu}}\email{a.bhambu@iitg.ac.in}

\author*[2]{\fnm{Arabin Kumar} \sur{Dey}}\email{arabin.k.dey@gmail.com}
\equalcont{These authors contributed equally to this work.}


\affil[1]{\orgdiv{Department}, \orgname{of Mathematics}, \orgaddress{\street{IIT Guwahati}, \city{Guwahati}, \postcode{781039}, \state{Assam}, \country{India}}}

\affil*[2]{\orgdiv{Department}, \orgname{of Mathematics}, \orgaddress{\street{IIT Guwahati}, \city{Guwahati}, \postcode{781039}, \state{Assam}, \country{India}}}



\abstract{In this paper we propose a novel procedure to construct a confidence interval for multivariate time series predictions using long short term memory network.  The construction uses a few novel block bootstrap techniques.  We also propose an innovative block length selection procedure for each of these schemes.  Two novel benchmarks help us to compare the construction of this confidence intervals by different bootstrap techniques.  We illustrate the whole construction through S\&P $500$ and Dow Jones Index datasets.}


%



\maketitle

\section{Introduction}\label{sec1}

    Time series forecasting has gained so much attraction due to its applications in the areas such as computational intelligence, statistical estimation, and finance \cite{pedrycz2013time}. Time series forecasting plays a vital role from an industrial point of view as it creates data-driven plans and makes wise business decisions. Machine learning algorithms are frequently used for time series forecasting. Any machine learning or forecasting technique forecasts the prediction with some degrees of random variations in the prediction for each observation. The point estimation of observation with some variation gives not much relevant information and can be misleading in some situations.  The general statistical approach resolves such problems by providing a confidence band around the predicted values. Recently, De et al. \cite{de2020construction} provided construction of such confidence intervals for univariate signals. No appropriate notion is available for constructing a confidence interval of multidimensional signal prediction through machine learning models.  

 We propose to solve this problem with the help of block bootstrap techniques. Radovanov and Marcikic \cite{radovanov2014comparison} performed a comparative study over non-overlapping block bootstrap, moving block bootstrap, stationary block bootstrap, and subsampling. Hongyi and Maddala \cite{hongyi1996bootstrapping} discussed the uses of bootstrap methods, recursive bootstraps, moving block bootstraps, and stationary bootstraps. None of the literature talks about the bootstrap in a transformational space. De et al. \cite{de2020construction} introduced a few novel block bootstraps in a transformed space. However, those constructions are available for univariate signals only. No study is available on block bootstrap in a multidimensional setup that comes under certain transformational space. In this paper, we extend further specific bootstrap algorithms in a high dimensional set up which is super effective in serving all purposes throughout the paper.
   
 A major problem in all constructions of block bootstrap methods in a multivariate time series signal is the choice of the optimal block length.  Demirel and Willemain (\cite{demirel2002generation}) use Higher-order crossings (HOC) to form the benchmark to find out the bootstrap length. This approach does not have a proper theoretical basis for non-stationary data and has many shortcomings. 
Recently, De et al. \cite{de2020construction} provided a heuristic approach to find the bootstrap block length which has two major shortcomings even though it works quite well in many non-stationary datasets : (1) it is for univariate time series (2) It does not have a solid theoretical foundation. The paper plans to extend the same concept in a multivariate set up which includes a two-fold transformation to propose a penalty-based empirical function to choose the optimal block length for bootstrap methods. 

 All constructions of confidence intervals include the minimization of confidence width given a choice of probabilistic confidence. However, extending the idea of minimization of confidence width to minimization of confidence area for signal confidence band is not enough for the construction of the best confidence interval, since it is also important to ensure that the confidence interval moves along the original time series signal, but includes the original series with a certain confidence level at each time step. We introduce two novel metrics which can serve this purpose in section \ref{exp}. 
 
  We organise the paper as follows. Different modified block bootstrap methods proposed in the multi-dimensional structure are available in Section 2. Section $3$ contains a brief discussion about the deep LSTM models. In Section $4$, we discuss the choice of block length in all above bootstrap procedures and proposed methodology for the construction of the confidence interval in higher dimensional setup. The dataset description, data pre-processing, the experimental setup are available along with proposed assessment metrics in Section $5$. Section 6 contains the results and related discussions. We conclude the paper in Section $7$. 
      
\section{Proposed Block Bootstraps for Multidimensional Time series}\label{sec2}

 In this section, we extend the notion of bootstrap described by \cite{de2020construction}.  The key difference of the usual bootstrap and the bootstrap described in that paper in univariate signal is it proposes to use a transformation of the dataset before going for bootstrap and retransform the bootstrap sample back to its original space.  The advantage of working in this set up is choice of optimal window-length selection becomes tractable in most of the datasets.  We propose to extend the same notion in higher dimensional system and observe all bootstrap procedure mimic the component signal well by these procedure. 
  
  We use the following steps to carry out the bootstrap. 

\begin{itemize}

\item First transform the full multivariate data into one or two fold transformational set up (e.g. log transform of ratios of the series and window scaling) for every component series.   

\item Perform usual block bootstrap sample on multivariate series dividing the time wise whole series into different blocks.

\item Return back to original sample space.

\end{itemize}       
  
\subsection{Description of each of the Block bootstraps}  
    
  The block bootstrap approach is one of the most popular bootstrap procedures in a time series domain as it preserves the dependent behavior of time-series within the block. The motivation behind proposing such a method is maintaining the time series behavior within the pseudo-sample because classical bootstrap methods resample the data, and then pseudo-samples are generated. The block bootstrap methods are introduced by Hall \cite{hall1995blocking}, and Carlstein \cite{carlstein1986use}. In block bootstrap methods, we divide the original time series signal into several blocks, and resample the data in each block using the classical i.i.d. bootstrap. 
  
\subsection{Non-Overlapping Block Bootstrap (NOBB)}

  Given the dataset,  $X_1, X_2, \cdots, X_n $, the non-overlapping block bootstrap method does not allow the overlapping of data segments contained in the sequences of blocks. Let us assume that, $l \in [1, n] $  is an integer, then the number of blocks $b \geq 1$ satisfies the relation, $b = \left[n/l\right]$, where $[.]$ is the greater integer function. The non-overlapping block bootstrap procedure for generating a new bootstrap sample is as follows:

\begin{enumerate}
	\setlength\itemsep{1em}
		
	\item Define, $I_{n,b} = \{1, l + 1, 2l + 1, \cdots, (b-1)l + 1\}$, if non-overlapping blocks are considered.
	
	\item Choose discrete independent random variables i.e., indices $i_1, i_2, \cdots, i_b$.
	
	\item The blocks are represented as $B_i = (X_{(i-1)l + 1}, \cdots, X_{il})$, where $1\leq i \leq k$ and are considered in end-to-end order and sampled together.
	
	\item Discard the last $(b-k)$ entries to form a bootstrap sample, $X_1^{\star}, X_2^{\star}, \cdots, X_m^{\star}$, where $m = lb$.
\end{enumerate}    

%

\subsection{Moving Block Bootstrap (MBB)}

  Let us consider the dataset $X_1, X_2, \cdots, X_n $, the moving block bootstrap tries to mimic the time-series behavior, $\hat{P_n} = F(A_n)$, where $A_n$ is the empirical function of the random sample distribution. Assume that $l \in [1, n] $ is an integer and holds the property that as $l \rightarrow \infty$ and $n \rightarrow \infty$, $l/n \rightarrow 0$. However, a detailed explanation for this method starts by assuming the block length, $l$, as a constant. If $B_i = (X_1, \cdots, X_{i+l-1}) $ represents the $i$-th block of time series with the block length, $l$, where $N = n - l + 1$ is the number of blocks in the pseudo-sample. The moving block bootstrap forms pseudo-samples by randomly selecting a certain number of blocks from the set ${B_1, B_2, \cdots, B_n}$. 
\par As a result, $B_1^{\star}, B_2^{\star}, \cdots, B_k^{\star}$ denotes a random sample with repetitions form the set ${B_1, B_2, \cdots, B_n}$, where each blocks have the same number of elements $l$. The representation of the block $B_i^{\star}$ presented as $(X_{(i-1)l + 1}, \cdots, X_{il})$, where $1\leq i \leq k$. Then, the bootstrap sample is $X_1^{\star}, X_2^{\star}, \cdots, X_m^{\star}$ with the block size $m = kl $ .

\subsection{Local Block Bootstrap (LBB)}

Given the dataset $X_1, X_2, \cdots, X_n $, then the procedure of generating a pseudo-sample from the local block bootstrap method is as follows:

\begin{enumerate}
	\setlength\itemsep{1em}
	
	\item Choose a block length, $l \in [1, n] $  is an integer, and a real number $B \in (0,1]$ such that $nB$ is an integer, where $l$ and $B$ are functions of $n$.
	
	\item For $m = 0, 1, \cdots, ([n/l] - 1)$, where $[.]$ is greatest integer function. Let $X^{\star}_{ml+j} = X_{i_m + j - 1}$ where, $1\leq j \leq l$ and $i_1, i_2, \cdots$ are independent random integer values satisfying the relation, $P(i_m = k) = W_{n,m}(k)$.
	
	\item The choice of $k$ is made by satisfying the relation $W_{n,m}(k) \sim Unif(J_{1,m}, J_{2,m})$ where $J_{1,m} = max\{1, ml - nB\}$ and $J_{2,m} = min\{n - l + 1, ml + nB\}$.
\end{enumerate}

\section{Long Short-Term Memory (LSTM)}

  Long short-term memory (LSTM) is a gating recurrent neural network (RNN) introduced to learn long-term dependencies of the time-series datasets \cite{hochreiter1997long}. The primary motivation of the network is to solve the gradient vanishing problem of RNNs with the help of the gating mechanism. There are three gates in each LSTM cell: input, forget, and output. The gates allow the network to selectively write information obtained from the last cell's output, selectively read information obtained from the intermediate step, and selectively forget irrelevant information. 

The mathematical operation of the network is as follows: 
\begin{equation*}
	\begin{aligned}
		\mathbf{i}_{t} &=\sigma\left(\mathbf{W}_{x i} \mathbf{x}_{t}+\mathbf{W}_{h i} \mathbf{h}_{t-1}+\mathbf{W}_{c i} \mathbf{c}_{t-1}+\mathbf{b}_{i}\right) \\
		\mathbf{f}_{t} &=\sigma\left(\mathbf{W}_{x f} \mathbf{x}_{t}+\mathbf{W}_{h f} \mathbf{h}_{t-1}+\mathbf{W}_{c f} \mathbf{c}_{t-1}+\mathbf{b}_{f}\right) \\
		\mathbf{c}_{t} &=\mathbf{f}_{t} \odot \mathbf{c}_{t-1}+\mathbf{i}_{t} \odot \tanh \left(\mathbf{W}_{x c} \mathbf{x}_{t}+\mathbf{W}_{h c} \mathbf{h}_{t-1}+\mathbf{b}_{c}\right) \\
		\mathbf{o}_{t} &=\sigma\left(\mathbf{W}_{x o} \mathbf{x}_{t}+\mathbf{W}_{h o} \mathbf{h}_{t-1}+\mathbf{W}_{c o} \mathbf{c}_{t}+\mathbf{b}_{o}\right) \\
		\mathbf{h}_{t} &=\mathbf{o}_{t} \odot \tanh \left(\mathbf{c}_{t}\right)
	\end{aligned}
\end{equation*}

\par where, $i_t, f_t, o_t, c_t, h_t$ are the input gate, forget gate, output gate, cell activation, and cell output activation values at time t, respectively. $\sigma$ is the sigmoid-activation and $\odot$ is the dot product between the two vectors. $W$ and $b$ stand for the weight matrices and bias vectors that connect various gates, respectively.

\section{Proposed Methodology}

   This research provides multiple contributions at different phases of the proposed solution.  We already demonstrated the proposition of forming all available bootstrap procedures in a high-dimensional structure through a specific transformed space.  One of the most challenging problems working with any bootstrap procedure is selection of the optimal block length.   Our proposed methodology successfully selects the optimal block length.  It seems the methodology is powerful enough to make its selection even in a very generic arbitrary high-dimensional time-series datasets.  We extend the construction of confidence intervals for each components of multidimensional signals.   We follow the architecture proposed through LSTM based models in our earlier contribution in univariate case for this purpose.   In later part of the paper, we propose a new additional benchmark to evaluate the best constructed confidence intervals by different bootstrap methods in this setup.   
   
\subsection{Choice of the optimal block length}

 The optimal choice of block length plays a crucial role in the block bootstrap methods used in the time-series aspect. Choosing the best block size is a difficult problem. Hall et al. \cite{hall1995blocking} state that the optimal block size mainly depends on the following factors: Auto-correlation structure, the length of the time-series signal, and the purpose of bootstrapping. Many authors also discussed the choices of the block length for estimating or determining the bootstrap mean or variances for dependent sets of data \cite{lahiri1999theoretical} \cite{politis2004automatic} \cite{nordman2014convergence}. 

\par Bootstrapping regenerates new samples by sampling the dataset with replacement. The primary motivation behind bootstrapping is to generate pseudo-samples that can mimic the time-series behavior. Any resampling algorithm is better if the generated pseudo-sample captures the variance of the original sample by minimizing the variation between the generated sample and the original sample. There are many methods that can measure this variance.   Quantification of this variance often uses construction of a distance function for the mean series from the blocks of the original and bootstrapped sample. Demirel and Williemain \cite{demirel2002generation} proposed Higher-order crossing (HOC) to find the best block length, but the method does not work for the non-stationary datasets. Later on, De et al. \cite{de2020construction} also introduced a methodology for univariate stock signals but did not give a specific explanation for the multivariate aspect.

\par This paper proposed a novel empirical distance function-based methodology to choose the block length, say $l$. The proposed methodology focuses on the variation between the filtered mean of the multivariate time series and the bootstrapped sample. We propose to calculate the following empirical distance function in a specific projected space, which consists of window scaling on the whole log-transform bootstrap series and the original log-ratio series:

$$ \frac{1}{M} \sum_{i=1}^{M} \frac{1}{T \times n_f} \sum_{j =1}^{n_f} \sum_{k=1}^{T} {(\tilde{X}^{\star}_{j, k}}^{(i)} - {\tilde{X}_{j, k}}^{(i)})^2  $$

where $T$ : look back number or time steps in each window  $n^{*}$: number of windows with length equals to look back number or time steps considered in LSTM model to make window scaling, $M$: = $\left[\frac{n^{*}}{l}\right]$ = number of blocks of length $l$ where $\left[ \frac{n^{*}}{l} \right]$ is the greatest integer function less than equals to $\frac{n^{*}}{l}$. and $n_f$: number of features in the multivariate time series, i.e., number of covariates in the input sample. Also, $ \tilde{X}^{\star (i)}_{j, k}$ and $\tilde{X}^{(i)}_{j, k}$ represent the filtered window-scaled mean up to the block $i$, the mean is calculated as:
\begin{align*}
	{\tilde{X}^{\star (i)}_{j, k}} &= \frac{1}{i \times l} \sum_{k^{'}}^{i \times l} X^{\star}_{j, k, k^{'}} \\
	\tilde{X}^{(i)}_{j, k} &= \frac{1}{i \times l} \sum_{k^{'} = 1}^{i \times l} X_{j, k, k^{'}}
\end{align*} 

where $\tilde{X}^{\star (i)}_{j, k, k^{'}}$ and $\tilde{X}^{(i)}_{j, k, k^{'}}$ are the bootstrap sample and original samples of $j$-th feature, $k$-th component of $k^{'}$-th window. Our intuition is to minimize the empirical function to get the optimal block length.

\par The above-proposed empirical function decreases with many oscillations with respect to the block length. The oscillations make selecting the optimal block length very difficult as oscillations depend on the autocorrelation of the time series dataset \cite{hall1995blocking}. The Convexity of the function depends on the appropriate selection of the series length if a penalty function is added to this empirical function. So, we have proposed a penalty-based empirical function that also took care of the convexity of the function, that is, 

$$ \frac{1}{M} \sum_{i=1}^{M} \frac{1}{T \times n_f}\sum_{j =1}^{n_f} \sum_{k=1}^{T} {(\tilde{X}^{\star}_{j, k}}^{(i)} - {\tilde{X}_{j, k}}^{(i)})^2 + l * \frac{log(n^{*})}{{n^{*}}^\alpha} $$

\begin{figure}[!h]
	\centering
		\includegraphics[width=0.9\textwidth]{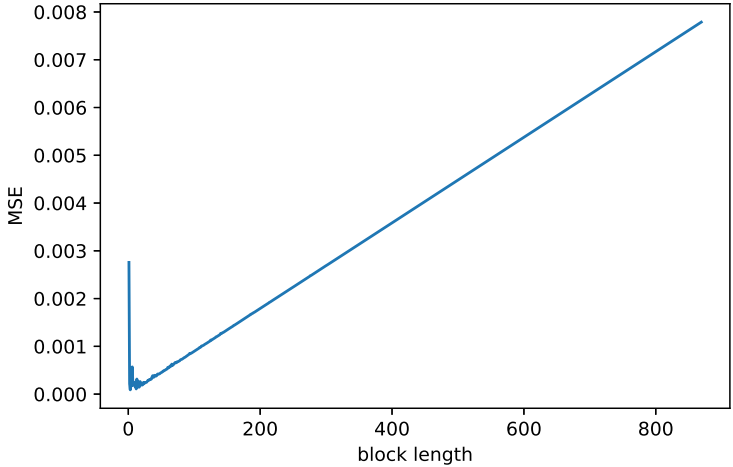}
		\caption{Optimal block length in non-overlapping block bootstrap for DJI}
		\label{nobb}
\end{figure}

\begin{figure}[!h]
	\centering
		\includegraphics[width=0.9\textwidth]{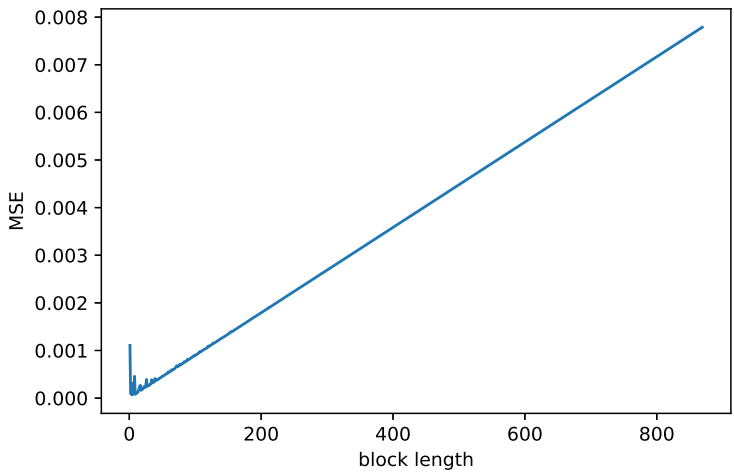}
		\caption{Optimal block length in moving block bootstrap for DJI}
		\label{mbb}
\end{figure}

\begin{figure}[!h]
	\centering
		\includegraphics[width=0.9\textwidth]{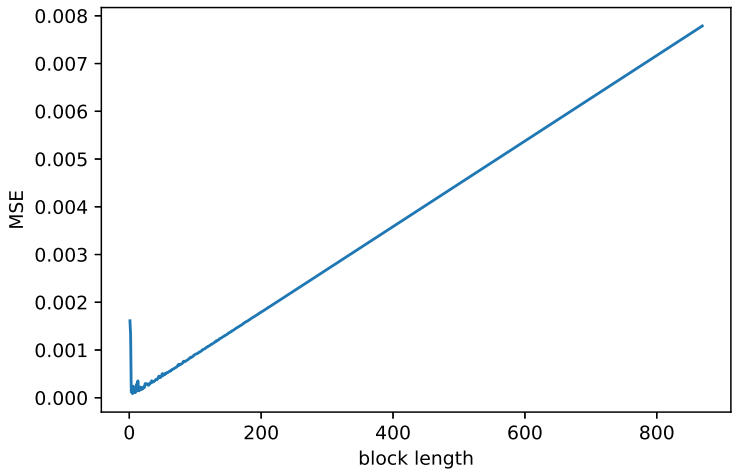}
		\caption{Optimal block length in local block bootstrap for DJI}
		\label{lbb}
\end{figure}

where, $ {\tilde{X}^{\star}_{j, k}} $ $ {\tilde{X}_{j, k}} $ are the filtered mean for bootstrap sample and log-transformed original sample. The above penalty-based empirical function improves the oscillation nature of the chosen convex function with respect to different window-lengths.  The log transformation of the sample observations is essential for the proposed approach, since it radically improves the structure of the objective function.  Therefore, we first log-transform the data and find the optimal block length.   We can then use any of the above bootstrap techniques to generate a bootstrap sample.  However,  these samples are not samples from the original sample space.   Therefore,  we reverse-transform these bootstrap samples before passing to any model.  Also, we need to choose suitably empirical function parameter $\alpha$ for better convergence.   We observe that empirical penalized mean square error behaves like a convex function for $\alpha = 2$ across all chosen our datasets.  We need better theoretical justification for these choice which we consider in a separate article.

The Figure-\ref{nobb} Figure-\ref{mbb} and Figure-\ref{lbb} correspond to each bootstrap methods for the DJI dataset, showing the block length with respect to the proposed penalty-based empirical distance function.  All figures provide a supportive evidence for the fact that the above mentioned approach successfully achieves the goal of choosing the optimal block length.   The procedure works even for a highly volatile, and non-stationary form of data. A more robust theoretical analysis for the proposed approach require separate attention.

\subsection{Construction of confidence interval for multivariate setup}

  In this section, we provide the construction of confidence interval based on above generated bootstrap samples for a multivariate time-series using the LSTM network. The procedure for the confidence interval is as follows:

\begin{enumerate}
	\setlength\itemsep{0.5em}
	
	\item Generate bootstrap samples that mimics the input's original sample with low variation.
	\item For each bootstrap sample, fit an LSTM model and predict on the test dataset.
	\item Calculate the 95\% non-parametric confidence interval at each time step for each feature component separately from the test dataset.
	
\end{enumerate}

   The key Construction steps of the confidence interval for univariate and multivariate signals are almost identical.  However, each step requires special attention due to its inherent high-dimensional complexity.  In this paper, we construct the multidimensional confidence interval consisting of the following component variables, Open price, High price, Low price, Closing price, and Volume.  First, we select the optimal block length for this time series signal which requires the data in log-transformed space.  Final bootstrap samples on original data again demand a reverse transformation.  We pass the bootstrap train samples to the multivariate LSTM network and get the multivariate prediction on the test dataset.   Finally, we calculate a 95\% non-parametric confidence interval at each time step based on all available signal predictions at a particular time point.
   
\section{Experiments}
\label{exp}

\subsection{Data description and partitioning}

\begin{figure}[!ht]
	\centering
	 \includegraphics[width=0.49\textwidth]{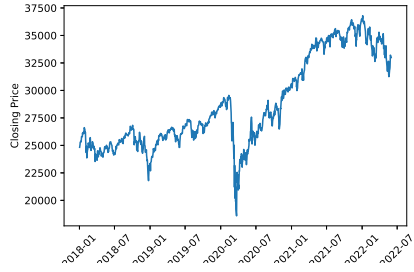}
	
	\caption{Visualization of the time series dataset (a) DJI index dataset }
	\label{fig:data1}
\end{figure}

\begin{figure}[!ht]
	\centering
	 \includegraphics[width=0.49\textwidth]{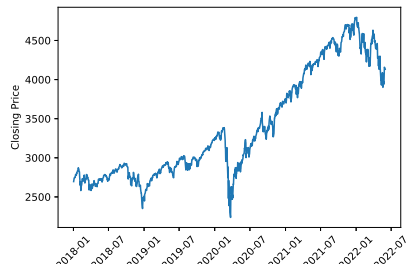}
	
	\caption{Visualization of the time series dataset (b) S\&P $500$ dataset \label{fig:data2}}
	
\end{figure}

  The dataset comprises two datasets: Dow Jones Industrial (DJI) Average is a stock index of $30$ leading US companies, and S\&P $500$ is a stock index that tracks the performance of more than $500$ US companies from $1$st Jan 2018 to $1$st July 2022. The sample size for both datasets is $1111$.  All time-series data are available in Figure-\ref{fig:data1} and Figure-\ref{fig:data2}.
\par
\begin{figure}[!ht]
	\centering
	\includegraphics[width=0.49\textwidth]{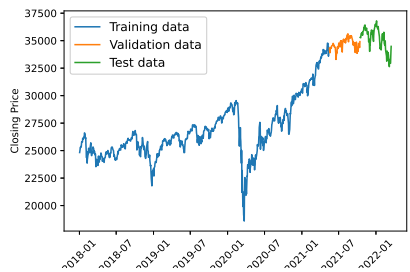}

	\caption{Train-Validation-test splitting of the time series dataset (a) DJI index dataset \label{fig:splitdata1}}
	
\end{figure}

\begin{figure}[!ht]
	\centering
	\includegraphics[width=0.49\textwidth]{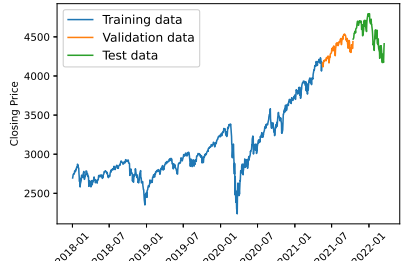} 

	\caption{Train-Validation-test splitting of the time series dataset (b) S\&P $500$ dataset \label{fig:splitdata2}}
	
\end{figure}

 The dataset has three parts: Training, Validation, and test datasets, see Figure-\ref{fig:splitdata1} and Figure-\ref{fig:splitdata2}. We have treated the first $80\%$ data as training, later $10\%$ as validation, and the last $10\%$ as test data. The optimal block length is calculated over the bootstrapped train data and validated on the validation dataset. After passing the bootstrapped data to the model, we use the validation dataset for tuning the hyper-parameters of the model setup.  We use the confidence interval using the test data.

\subsection{Data-preprocessing}

The data is scaled to the range between $0$ to $1$ to stabilize the training of the deep LSTM model. 

$$ x_{std} (i) = \frac{x_i - min(x)}{max(x) - min(y)} $$

\par We transform the scaled data into log-scaled data and generate the pseudo-samples with the help of the block bootstrap method. The log transformation helps the data to behave more like a normal distribution. We observe that the chosen transformation also pulls each of the components towards stationarity which makes the penalized filtrated squared difference of the means in a window-scaled projected space, a stable convex function of the block length. We resample the data through different bootstrap methods after choosing the block length. An inverse transform on these bootstrapped samples on a transformed space takes the data back to its original space mimicking the original high dimensional data. We divide each bootstrapped sample into train, test, and validation sets to use in LSTM or other model-framework.

\subsection{Computing specification and software setup}

\par The CPU used in this work is Intel(R) Core i$5$-$7500$ CPU @ $3.40$Hz. All codes are available using python $3.9$.  We use the following set of modules : Pandas $1.4.3$, sklearn $1.1.1$, Keras $2.9.0$, and TensorFlow $2.9.1$.

\subsection{Assessment Metrics}

 The paper adapts three assessment metrics for deciding the best confidence interval. Let us define the measure MAD for calculating the mean absolute difference between original observations and average predictions of the predicted confidence band. Similarly, SABWD denotes the sum of the absolute difference between the lower and upper confidence band, and MSD calculates the mean squared difference between original observations and average predictions of the predicted confidence band. We prefer to use both MAD and SABWD as the benchmark to assess all confidence intervals for each component series. However, we can also use mean squared difference (MSD) instead of MAD.

\subsubsection{Mean Absolute Difference (MAD):}

  The benchmark MAD calculates the mean of the absolute variation between the average of the confidence band and the original values. Since the metric measures the variance between the average value of the predicted confidence interval and the original observation at every timestep,  a low value of MAD ensures that the predicted confidence interval moves along the original time-series. The mathematical expression of MAD for $j$th component series is as follows:

\begin{align}
	MAD_{j} &= \frac{1}{T} \sum_{i = 1}^{T} \mid y_{ij} - a_{ij} \mid \nonumber \\
\end{align}

Where $T$ is the length of the time series signal and $a_{ij} = \frac{U_{ij} + L_{ij}}{2}$ are the average predictions of the predicted confidence band.  $U_{ij}$ and $L_{ij}$ are the upper and lower bounds of $j$th component series at $t_{i}$th time stamp.

\subsubsection{Mean Squared Difference (MSD):}

  MSD is used to determine the mean of the squared difference of the original value and the average of the upper and lower bounds. This metric can be an alternative similar measure to MAD.  A low value of the metric also ensures that the predicted confidence interval moves along the direction of the original time series signal. The MSD for $j$-th component series is as follows : 

\begin{align*}
	MSD_{j} &= \frac{1}{T} \sum_{i = 1}^{T} (y_{ij} - a_{ij})^2
\end{align*}

\subsubsection{Absolute Band Width Difference (ABWD):}

  The above metrics ensure that the confidence interval goes along the original time-series.  However, these metrics can not figure out the patterns of the width of the confidence interval in the given time interval.  Therefore, we set another benchmark, which is the sum of the absolute value of confidence width at each time instance or ABWD. The measure provides a naive approximation of the whole or significant portion of the area covered by the confidence band.   The best confidence interval should possess the lowest possible area covered by the confidence band.  The ABWD value provides an essential comparing factor for different bootstrap confidence bands.  We can define the ABWD for $j$th component as:

\begin{align*}
	ABWDE_{j} &= \sum_{i = 1}^{T} \mid U_{ij} - L_{ij} \mid
\end{align*}

All notations follow usual convention. 

\subsection{Hyper-parameters for the deep LSTM model}

\par To obtain the best performance of the LSTM model, we consider the hyper-parameter tuning for the LSTM. The parameters tuned along with the search space are available in Table \ref{table:HP}. After obtaining the best hyper-parameters, we train the model and construct the confidence intervals based on the related predictions.

\begin{table}[!h]
	\caption{Hyper-parameter domain of the parameters for the LSTM model}
	\begin{center}
		\begin{tabular}{ c c }
			\hline
			\textbf{Parameters} & \textbf{Search Space}\\
			\hline
			Batch size & [8, 16, 32, 64]\\
			Hidden size & [16, 32, 64, 128] \\ 
			Number of Layers & [1, 2, 3, 4] \\  
			Learning rate & [0.0001, 0.001, 0.01] \\
			Dropout rate & [0.1, 0.2, 0.3, 0.4] \\
			Activation function & [`relu', `silu', `sigmoid', `tanh'] \\
			Optimizer & [`Adam', `RMSprop', `SGD', `Adagrad', `Adamax'] \\
			\hline
		\end{tabular}
	\end{center}
	\label{table:HP}
\end{table}

\section{Results}

\subsection{Performance evaluation}

   We calculate the proposed metrics to evaluate the confidence intervals for a multivariate signals based on the original data.   We can check MSD, MAD, and ABWD for each bootstrap method and for each component series, such as open price, high price, low price, closing price, and volume in Table-\ref{Table:DJI} and Table-\ref{Table:SP} respectively.  The Local block bootstrap (LBB) outperforms other bootstrap methods in our experiments.  The result is true for both DJI stock index data (see Table-\ref{Table:DJI}) and for S\&P 500 stock index, which is available in Table-\ref{Table:SP}.  
   
\begin{table}[!h]
	\caption{The MSD, MAD, and ABWD for DJI stock index dataset. The best bootstrapping technique with the least MSD, MAD, and ABWD is marked in bold. }
	\label{Table:DJI}
	\begin{center}
		\begin{tabular}{ccccc}
			\hline
			& Bootstrap Methods & MSD & MAD & ABWD\\
			\hline
			
			\multirow{3}{*}{Open Price}& 
			$\quad$NOBB$\quad$  & $423577.78$ & $521.95$ & $102842.9$ \\
			& $\quad$MBB$\quad$ & $122350.40$ & $269.28$ & $70829.4$ \\
			& $\quad$LBB$\quad$ & $\textbf{69056.29}$ & $\textbf{199.93}$ & $\textbf{62479.8}$ \\
			
			\hline \multirow{3}{*}{High Price} & $\quad$NOBB$\quad$  & $400258.80$ & $511.40$ & $102246.5$ \\
			& $\quad$MBB$\quad$ & $142436.70$ & $294.90$ & $77182.3$ \\
			& $\quad$LBB$\quad$ & $\textbf{99061.31}$ & $\textbf{244.62}$ & $\textbf{65184.8}$ \\
			
			\hline\multirow{3}{*}{Low Price}& 
			$\quad$NOBB$\quad$  & $520594.69$ & $580.24$ & $108897.8 $ \\
			& $\quad$MBB$\quad$ & $199595.09$ & $341.32$ & $87625.0$ \\
			& $\quad$LBB$\quad$ & $\textbf{136009.59}$ & $\textbf{301.05}$ & $\textbf{67656.0}$ \\
			
			\hline \multirow{3}{*}{Closing Price} & $\quad$NOBB$\quad$  & $562463.66$ & $579.19$ & $104487.9$ \\
			& $\quad$MBB$\quad$ & $283476.60$ & $431.77$ & $70167.3$ \\
			& $\quad$LBB$\quad$ & $\textbf{222248.98}$ & $\textbf{381.74}$ & $\textbf{64935.7}$ \\
			
			\hline \multirow{3}{*}{Volume}
			& $\quad$NOBB$\quad$  & $8.56e+15\quad$ & $6.63e+07\quad$ & $7.99e+09\quad$ \\
			& $\quad$MBB$\quad$ & $6.62e+15\quad$ & $5.58e+07\quad$ & $7.47e+09\quad$ \\
			& $\quad$LBB$\quad$ & $\textbf{6.27e+15}\quad$ & $\textbf{5.52e+07}\quad$ & $\textbf{6.95e+09}\quad$ \\
			
			\hline
		\end{tabular}
	\end{center}
\end{table}

\begin{table}[!h]
	\caption{The MSD, MAD, and ABWD for S\&P $500$ stock index dataset. The best bootstrapping technique with the least MSD, MAD, and ABWD is marked in bold.}
	\label{Table:SP}
	\begin{center}
		\begin{tabular}{ccccc}
			\hline
			& Bootstrap Methods & MSD & MAD & ABWD\\
			\hline
			
			\multirow{3}{*}{Open Price}& 
			$\quad$NOBB$\quad$  & $6946.36$ & $65.72$ & $18181.5$ \\
			& $\quad$MBB$\quad$ & $4118.35$ & $48.47$ & $12993.9$ \\
			& $\quad$LBB$\quad$ & $\textbf{2209.72}$ & $\textbf{35.88}$ & $\textbf{11545.4}$ \\
			
			\hline \multirow{3}{*}{High Price} & $\quad$NOBB$\quad$  & $6752.75$ & $65.76$ & $17270.2$ \\
			& $\quad$MBB$\quad$ & $4073.18$ & $50.24$ & $12771.3$ \\
			& $\quad$LBB$\quad$ & $\textbf{2788.50}$ & $\textbf{42.34}$ & $\textbf{11857.1}$ \\
			
			\hline \multirow{3}{*}{Low Price}& 
			$\quad$NOBB$\quad$  & $8383.30$ & $73.70$ & $16886.9$ \\
			& $\quad$MBB$\quad$ & $5736.40$ & $58.16$ & $15095.6$ \\
			& $\quad$LBB$\quad$ & $\textbf{4027.22}$ & $\textbf{50.35}$ & $\textbf{11601.9}$ \\
			
			\hline \multirow{3}{*}{Closing Price} & $\quad$NOBB$\quad$  & $10198.64$ & $80.64$ & $15953.9$ \\
			& $\quad$MBB$\quad$ & $8053.63$ & $72.07$ & $12183.8$ \\
			& $\quad$LBB$\quad$ & $\textbf{6342.93}$ & $\textbf{65.40}$ & $\textbf{10855.8}$ \\
			
			\hline \multirow{3}{*}{Volume} & 
			$\quad$NOBB$\quad$  & $4.33e+17\quad$ & $\textbf{4.36e+08}\quad$ & $6.82e+10$ \\
			& $\quad$MBB$\quad$ & $4.41e+17\quad$ & $4.47e+08\quad$ & $6.75e+10$ \\
			& $\quad$LBB$\quad$ & $\textbf{6.50e+10}\quad$ & $3.83e+17\quad$ & $\textbf{4.25e+08}$ \\
			
			\hline
		\end{tabular}
	\end{center}
\end{table}

\subsection{Graphical View of the Confidence bands}

\begin{figure}[!ht]
	\centering
	\includegraphics[clip,width=0.7\columnwidth]{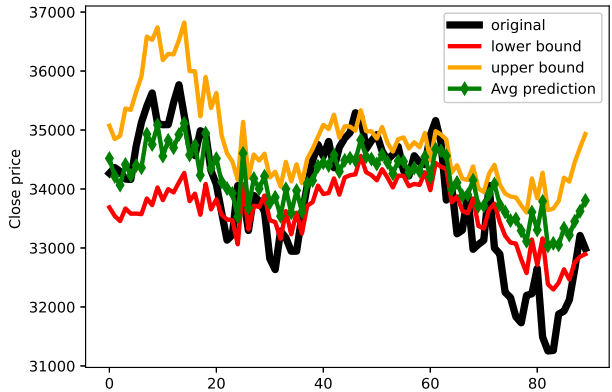} 
	\caption{(a) Confidence interval for Closing price obtained through NBB on LSTM using DJI dataset \label{fig:DJI11}}
\end{figure}

\begin{figure}[!ht]	
	\includegraphics[clip,width=0.7\columnwidth]{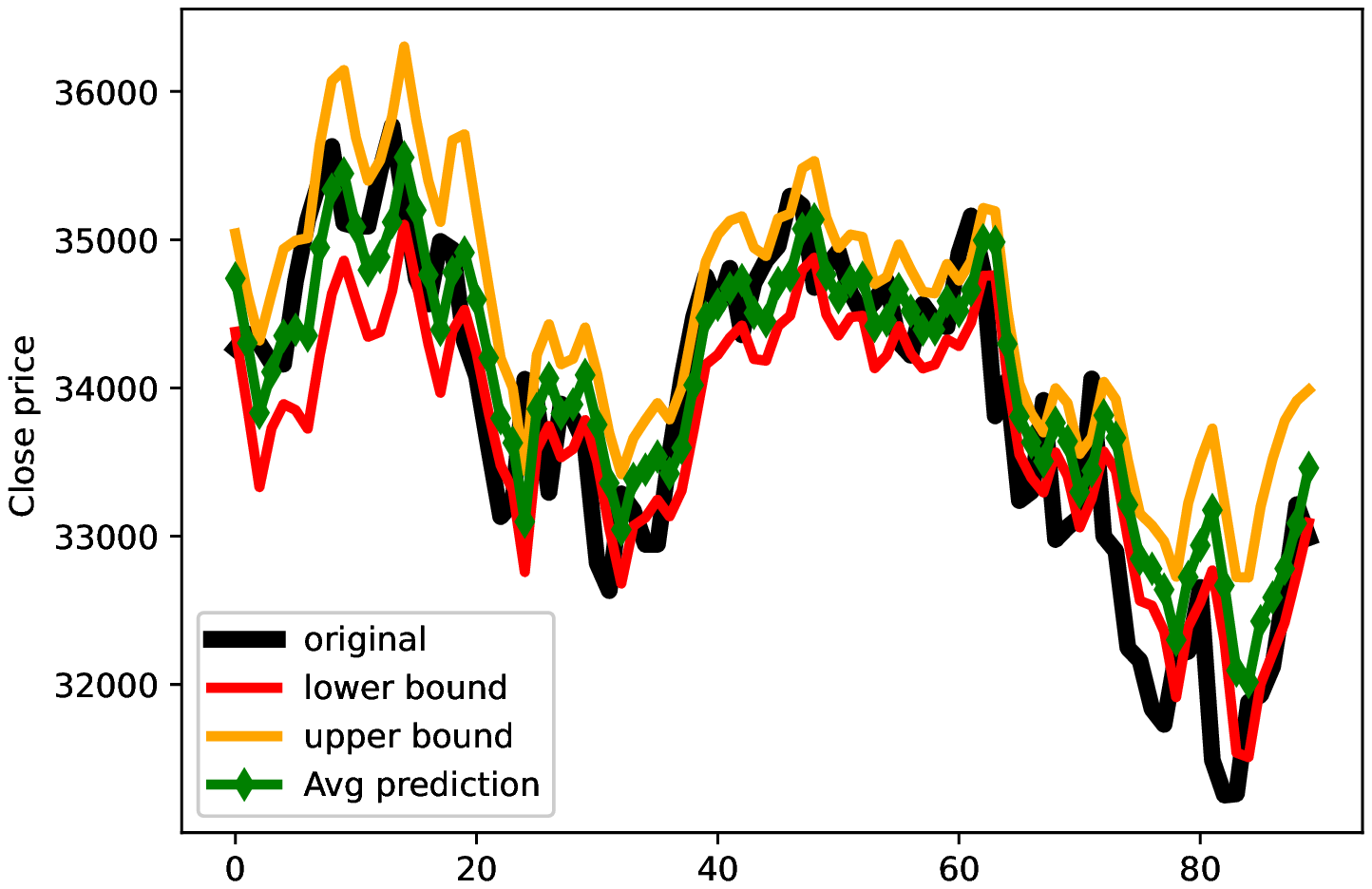}
	\caption{(b) Confidence interval for Closing price obtained through MBB on LSTM using DJI dataset 	\label{fig:DJI21}}
\end{figure}

\begin{figure}[!ht]	
	\includegraphics[clip,width=0.7\columnwidth]{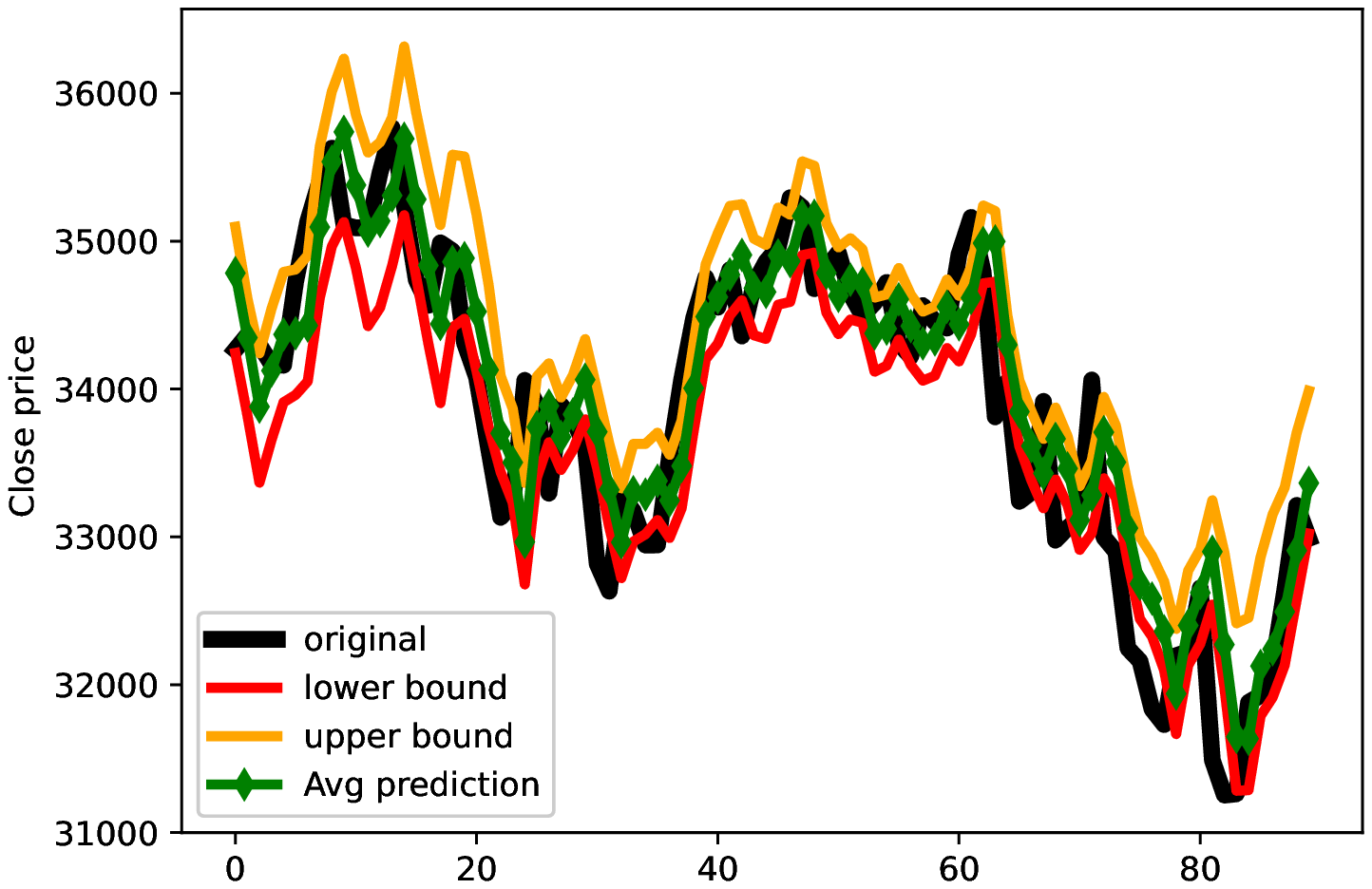}
	\caption{(c) Confidence interval for Closing price obtained through LBB on LSTM using DJI dataset \label{fig:DJI31}}
	
\end{figure}

\begin{figure}[!ht]
	\centering
	\includegraphics[clip,width=0.7\columnwidth]{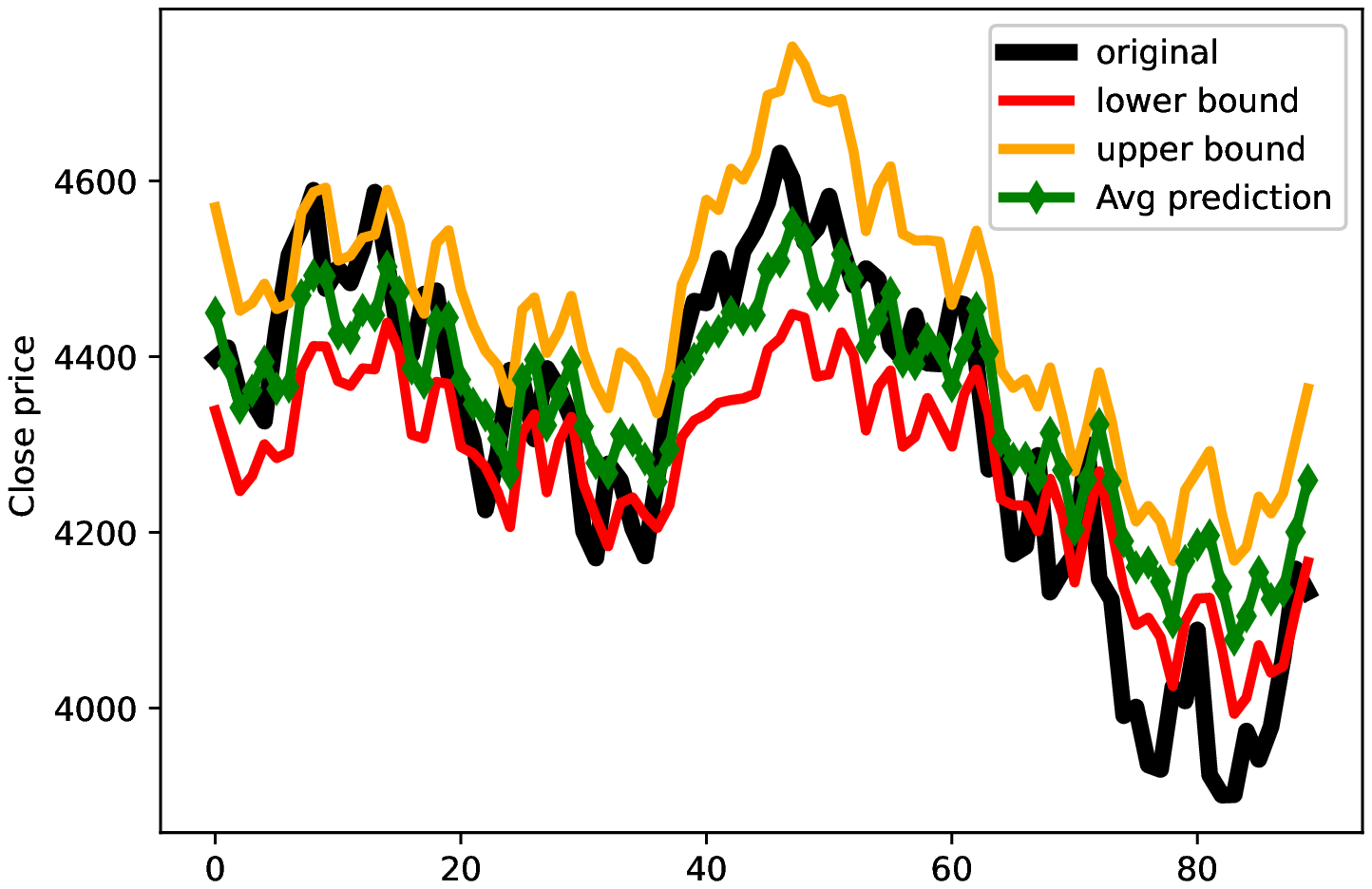}
  	\caption{(a) Confidence interval for Closing price obtained through NBB on LSTM using S\&P $500$ dataset \label{fig:SP12}}
\end{figure}

\begin{figure}[!ht]
	\centering
	\includegraphics[clip,width=0.7\columnwidth]{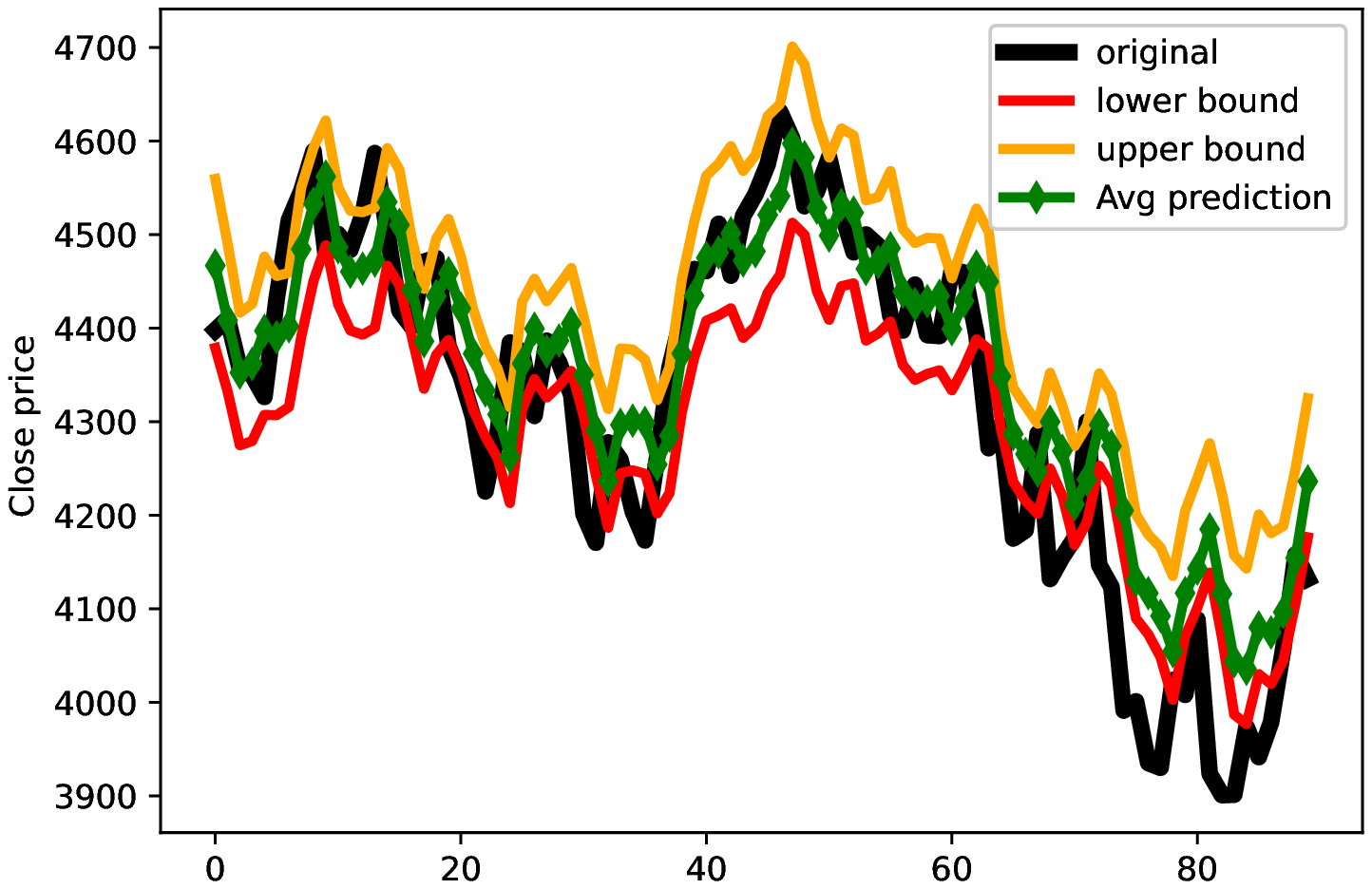} 
	\caption{(b) Confidence interval for Closing price obtained through MBB on LSTM using S\&P $500$ dataset \label{fig:SP22}}
\end{figure}

\begin{figure}[!ht]
	\centering
	\includegraphics[clip,width=0.7\columnwidth]{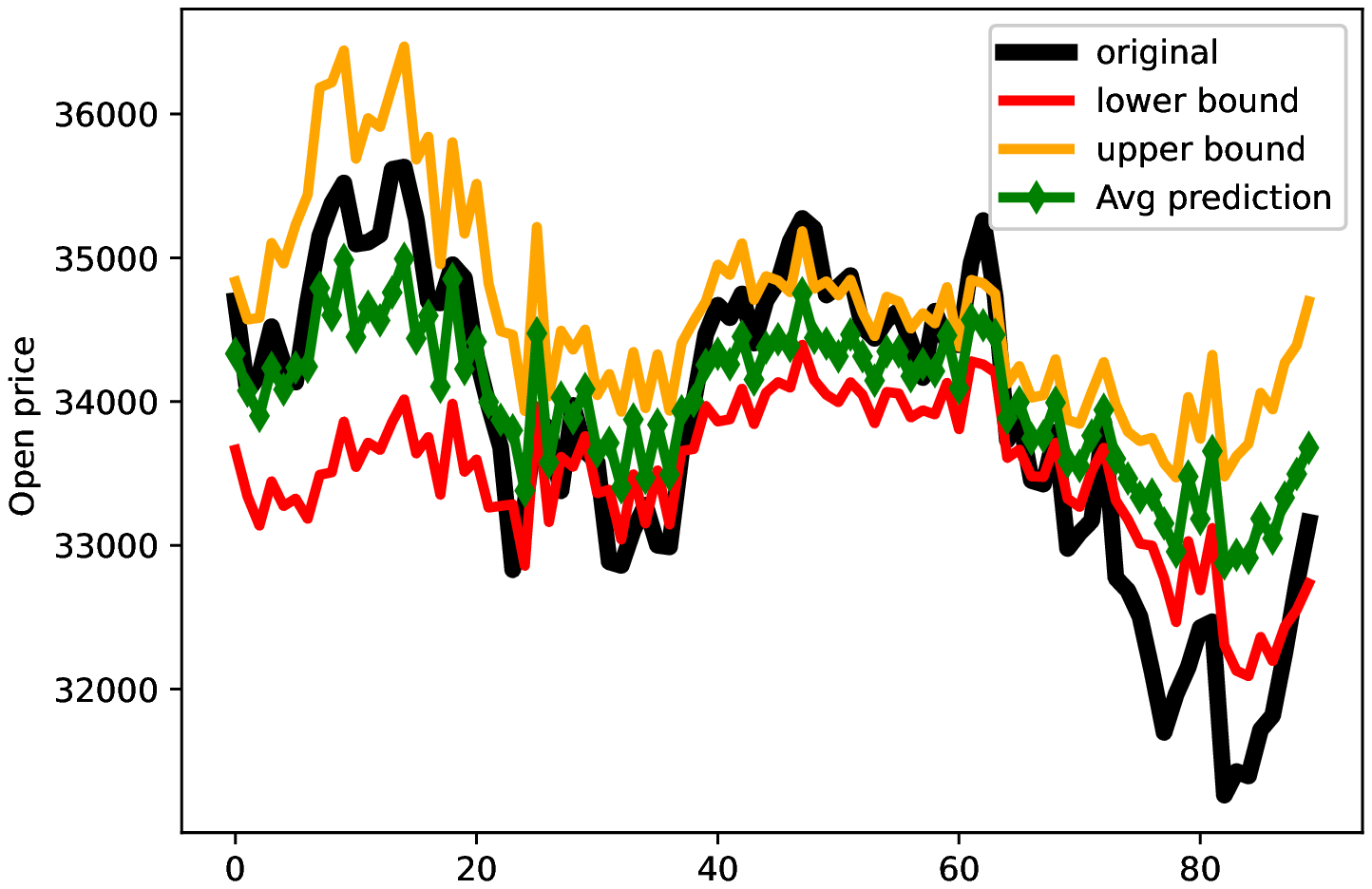}
	\caption{(a) Confidence interval for Open price obtained through NBB on LSTM using DJI dataset \label{fig:DJInbb}}
\end{figure}

\begin{figure}[!ht]	
	\centering 
	\includegraphics[clip,width=0.7\columnwidth]{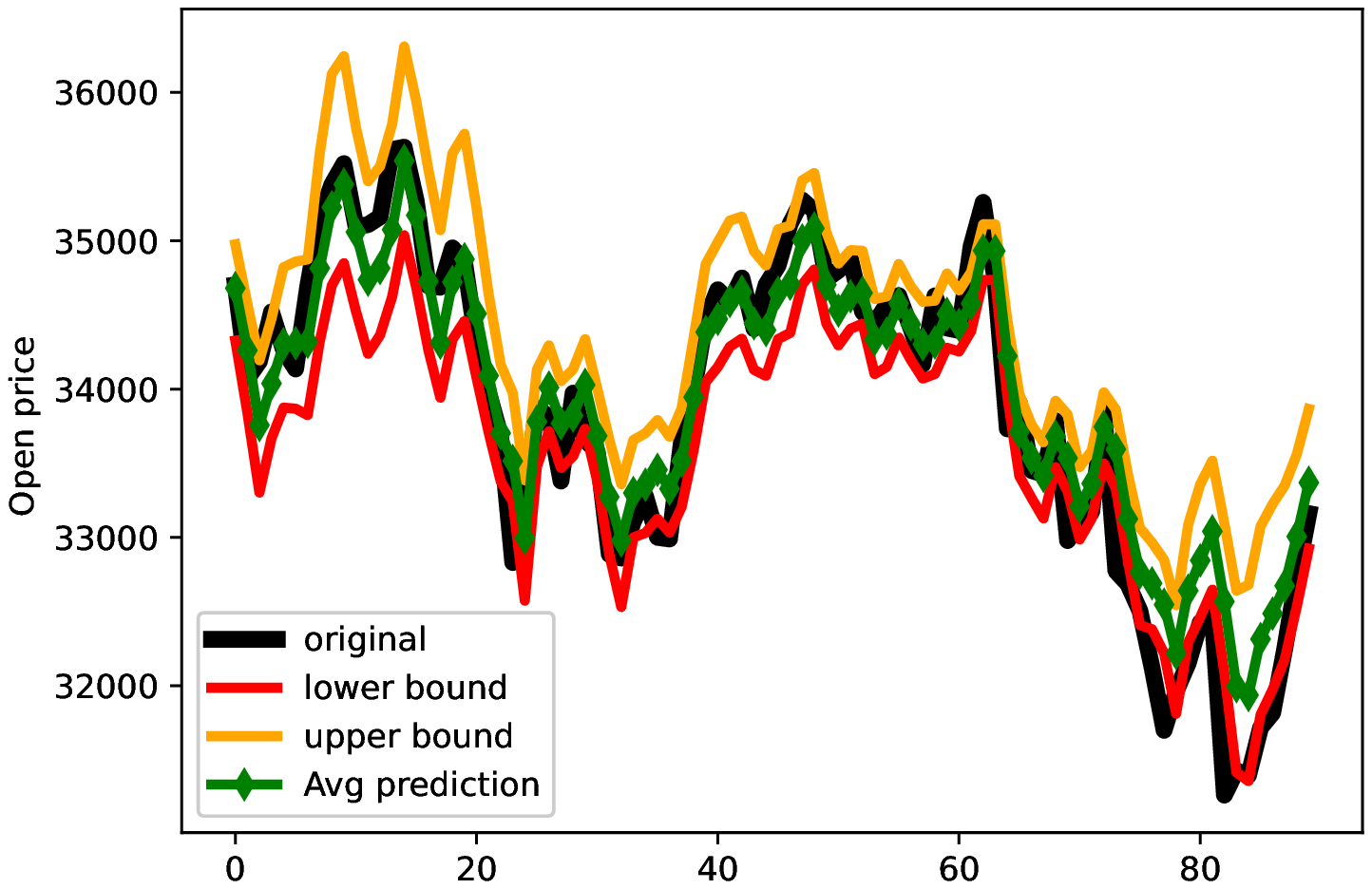} 
	\caption{(b) Confidence interval for Open price obtained through MBB on LSTM using DJI dataset \label{fig:DJImbb}}
\end{figure}

\begin{figure}[!ht]
	\centering	
	\includegraphics[clip,width=0.7\columnwidth]{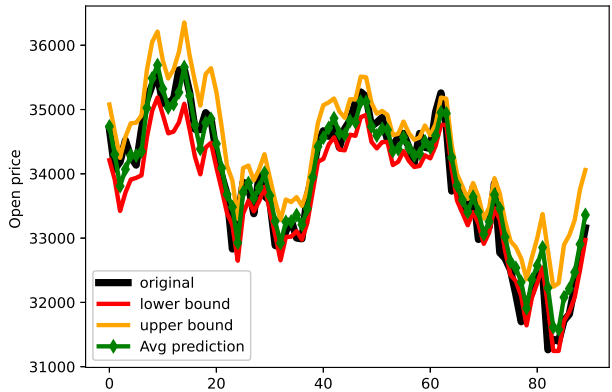}
	\caption{(c) Confidence interval for Open obtained through LBB on LSTM using DJI dataset \label{fig:DJIlbb}}
\end{figure}

\begin{figure}[!ht]
 	\centering 
	\includegraphics[clip,width=0.7\columnwidth]{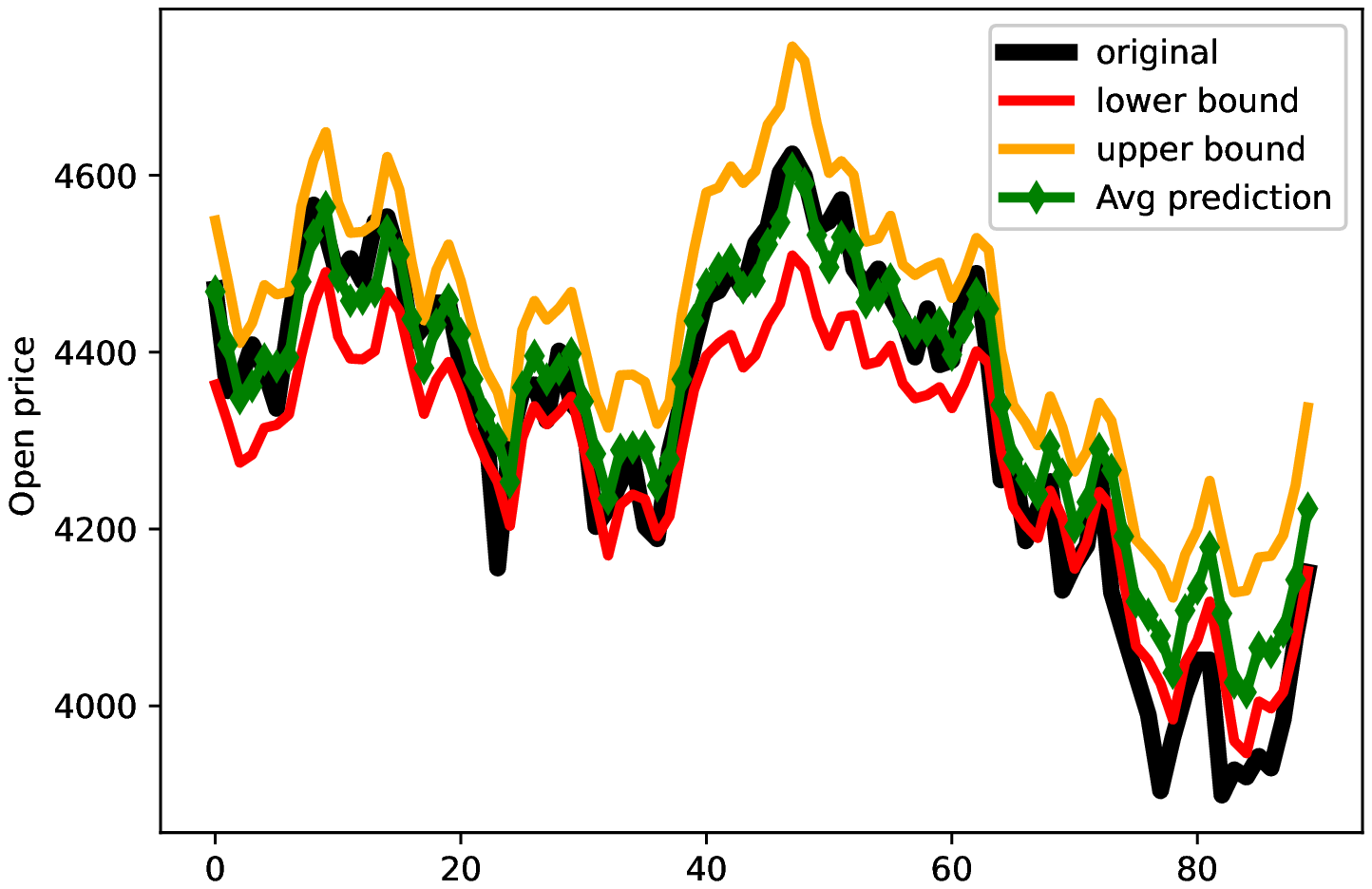}
	\label{fig:SP}
	\caption{(b) Confidence interval for Open price obtained through MBB on LSTM using S\&P $500$ dataset \label{fig:SPImbb}}
\end{figure}

\begin{figure}[!ht]	
 	\centering  
	\includegraphics[clip,width=0.7\columnwidth]{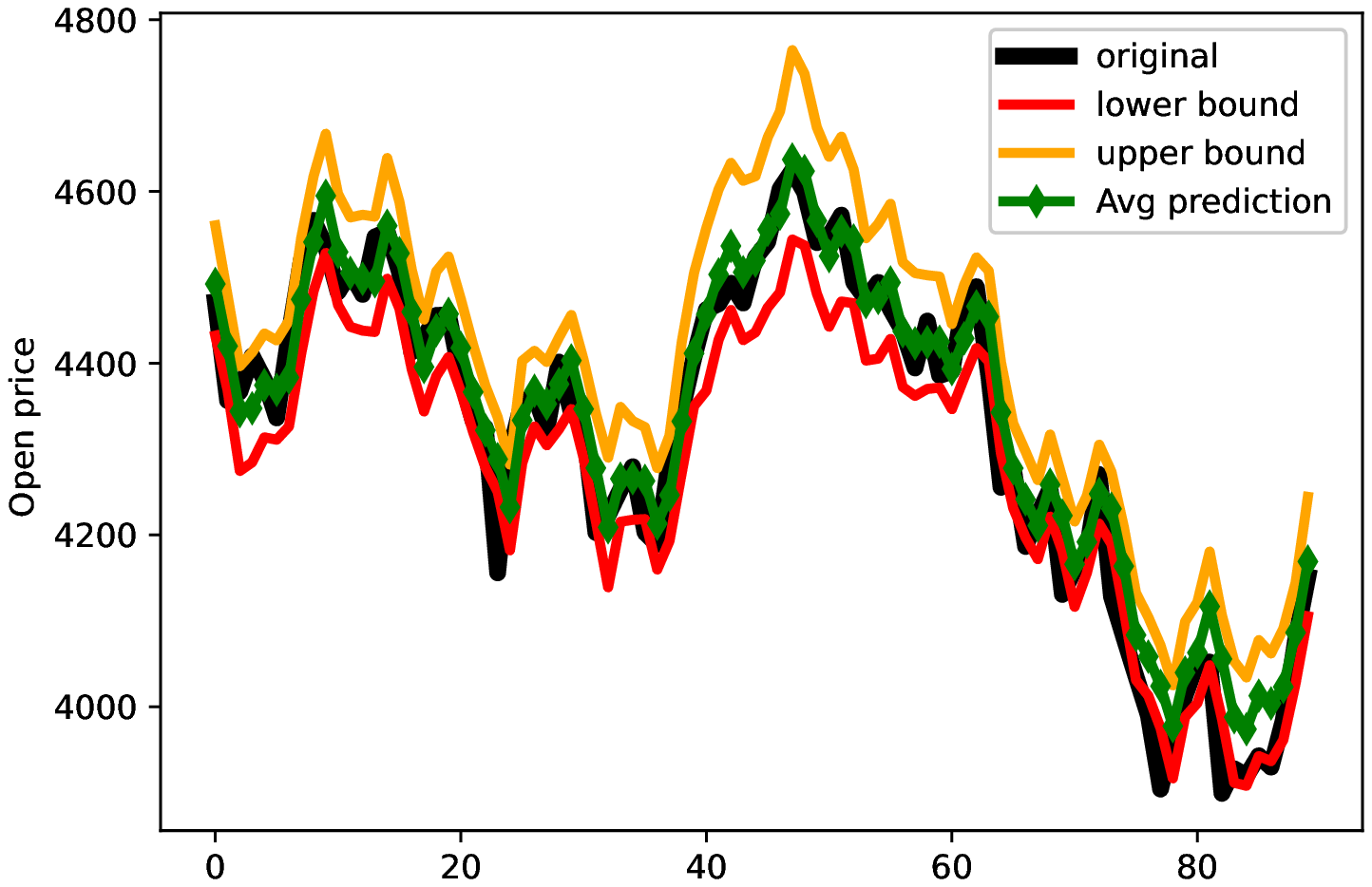}
	\caption{(c) Confidence interval for Open obtained through LBB on LSTM using S\&P $500$ dataset \label{fig:SPIlbb}}
\end{figure}

   We show the graphical view of confidence bands only for two components (closing price and opening price) out of the simultaneous five components series prediction.  Figure \ref{fig:DJI11}, Figure \ref{fig:DJI21} and Figure \ref{fig:DJI31} represent the confidence intervals for the closing price for the DJI through NBB, MBB, and LBB, respectively, whereas Figures \ref{fig:SP12}, Figure \ref{fig:SP22} and Figure \ref{fig:DJI32} shows confidence bands obtained by NBB, MBB, and LBB respectively for S\&P $500$ stock index datasets. The constructed confidence interval ensures to include original time series dataset with 95\% probability at every timestep. Moreover, figures have also information about the mean value of the upper and lower bound of the confidence interval and avg prediction. The average value shows the mean nature of the predicted confidence interval. Also, all consecutive figures from Figure \ref{fig:DJInbb} to Figure \ref{fig:SPIlbb} show the constructed confidence interval for the open price for the DJI and S\&P $500$ stock index datasets. We observe that the LSTM model with these bootstraps provides significantly good results, i.e., mean values of the chosen bootstrap methods are along the direction of the original time series. Since the approximate area of the confidence interval is lowest in LBB, and its mean value provides the best pictorial view in terms of the direction of the original sample, the proposed construction provides the best structure in LBB than other bootstraps.

\begin{figure}[!ht]
	\centering	
	\includegraphics[clip,width=0.7\columnwidth]{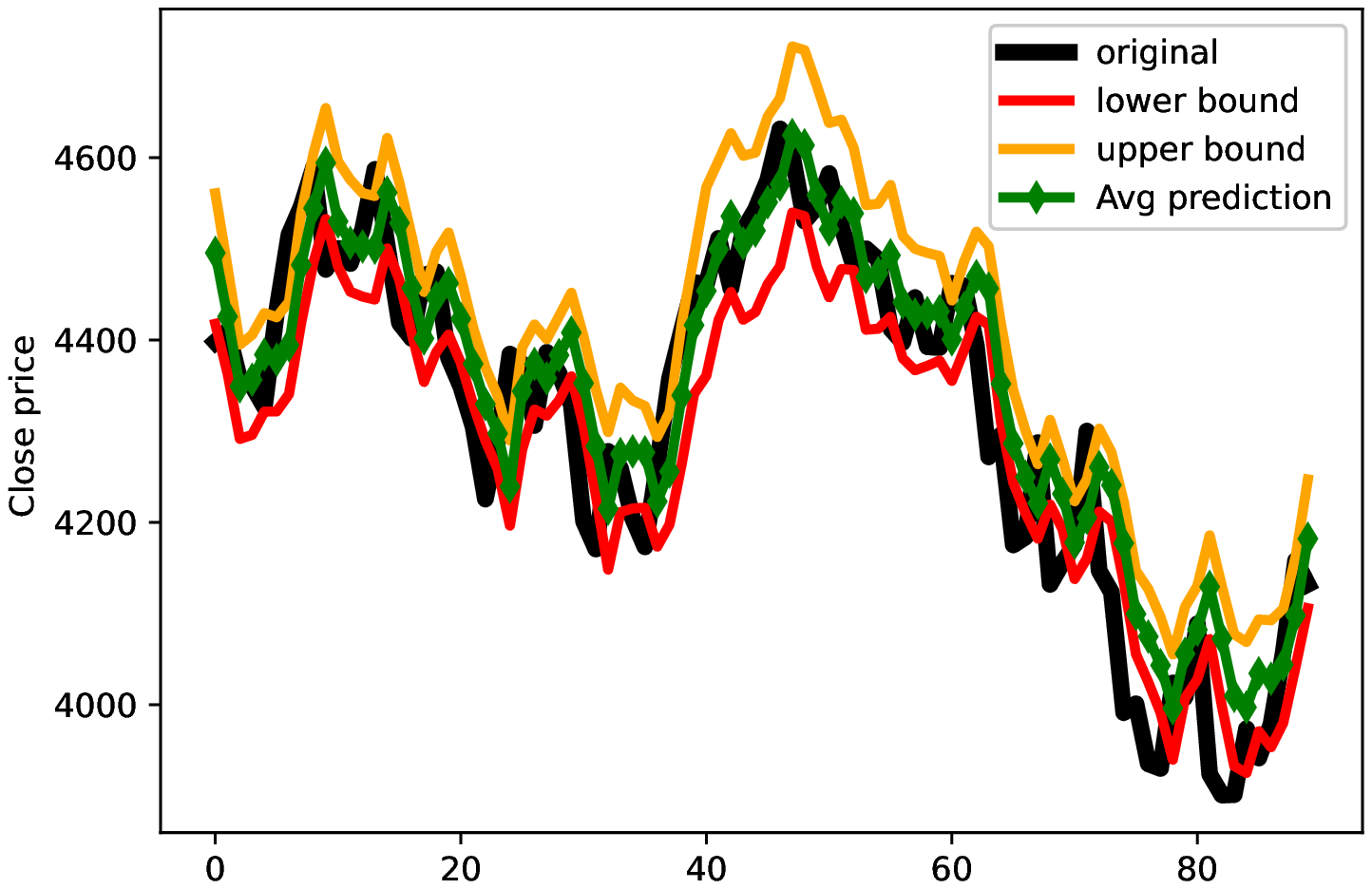}
	\caption{(c) Confidence interval for Closing price obtained through LBB on LSTM using S\&P $500$ dataset \label{fig:SP32}}
\end{figure}

\section{Conclusion}    The paper describe a specific construction of confidence interval for arbitrary multivariate signals through different bootstrap techniques. The confidence band construction depends mainly on resampling techniques, Neural network models used to predict multivariate-signals and statistical approach to form a confidence intervals at each time steps.  The proposed architecture in multivariate signals successfully provides an interval estimates of the overall prediction band.  The architecture includes a new scheme for different block bootstrap in higher dimensional set up, innovative block length selection and a novel proposition of benchmark to compare the construction techniques. The procedure needs more exploration in couple of directions, e.g., a comparison study of the constructions when we use better ensemble models, a more robust theoretical foundation of the proposed block length selection in terms of the order selection with respect to the sample size etc.  The work is in progress.        

\backmatter




\section*{Declarations}

\begin{itemize}
\item Funding :  The work is not related to any funding.

\item Conflict of interest/Competing interests :  There is no conflict of interest or competing interests for this work.

\item Ethics approval :  The paper does not have any ethics related issues.
 
\item Consent to participate : All co-authors have consent in participating for this research article. 

\item Consent for publication :  All co-authors have consent for keeping their name in related publication.

\item Availability of data and materials  :  All datasets are publicly available and downloaded directly from Yahoo finance.

\item Code availability :  The codes are available to each author.  It can be made publicly available through github on request.
 
\item Authors' contributions :  All authors have equal contribution. Aryan Bhambu is first author of the paper.  Dr Arabin Kumar Dey is the second and corresponding author.
\end{itemize}


\bibliography{dmk}

\end{document}